\documentstyle[epsfig,12pt,preprint,tighten,aps]{revtex}
\begin{document}

\draft

\begin{titlepage}
\rightline{February 2000}
\rightline{UM-P-012-2000}
\centerline{To appear in Physics Letters B}
\vskip 2cm
\centerline{\large \bf  
Can the mirror world explain the ortho-positronium lifetime
puzzle?}
\vskip 1.1cm
\centerline{R. Foot$^a$\footnote{Foot@physics.unimelb.edu.au} and 
S. N. Gninenko$^b$\footnote{Sergei.Gninenko@cern.ch}}
\vskip .7cm
\centerline{{\it $^a$ Research Centre for High Energy Physics}}
\centerline{{\it School of Physics}}
\centerline{{\it University of Melbourne}}
\centerline{{\it Parkville 3052 Australia}}
\centerline{{\it $^b$ CERN, CH-1211 Geneva 23, Switzerland}}
\vskip 2cm

\centerline{Abstract}
\vskip 1cm
\noindent
We suggest that the discrepant lifetime measurements of
ortho-positronium can be explained by ortho-positronium
oscillations into mirror ortho-positronium. 
This explanation can be tested in future vacuum experiments.

\end{titlepage}
A ``mirror universe'' is predicted to exist if parity and/or
time reversal are unbroken symmetries of nature\cite{ly,flv}.
The idea is 
that for each ordinary particle, such as the photon, electron, proton
and neutron, there is a corresponding mirror particle, of 
exactly the same mass as the ordinary particle. 
The parity symmetry interchanges the ordinary particles with the
mirror particles so that the properties of the mirror
particles completely mirror those of the ordinary particles.
For example the mirror proton and mirror electron are stable and 
interact with the mirror photon in the same way in which the
ordinary proton and electron interacts with the ordinary photons.
The mirror particles are not produced
in Laboratory experiments just because they couple very
weakly to the ordinary particles. In the modern language of gauge
theories, the mirror particles are all singlets under 
the standard $G \equiv SU(3)\otimes SU(2)_L \otimes U(1)_Y$
gauge interactions. Instead the mirror
particles interact with a set of mirror gauge particles,
so that the gauge symmetry of the theory is doubled,
i.e. $G \otimes G$ (the ordinary particles are, of 
course, singlets under the mirror gauge symmetry)\cite{flv}.
Parity is conserved because the mirror particles experience
$V+A$ (i.e. right-handed) mirror weak interactions
while the ordinary particles experience the usual $V-A$ (i.e.
left-handed) weak interactions.  Ordinary and mirror
particles interact with each other predominately by
gravity only.
At the present time there is some experimental evidence
that mirror matter exists coming from cosmology\cite{dark} as well as from
the neutrino physics anomalies\cite{neu}.

It was realized some time ago by Glashow\cite{gl} that
the orthopositronium system provides one sensitive
way to search for the mirror universe. 
The idea is that small kinetic mixing of the ordinary and mirror
photons may exist which would mix ordinary and mirror 
orthopositronium, leading to maximal orthopositronium -
mirror orthopositronium oscillations. 

The ground state of orthopositronium (o-Ps) decays predominately into
3 photons with a theoretical decay rate computed to be
(see e.g. \cite{ck} for a review)
\begin{eqnarray}
\Gamma  &=&
{2(\pi^2 - 9)\alpha^6 m_e \over 9\pi}\left[
1 - 10.28661{\alpha\over \pi} - {\alpha^2 \over 3}ln{1 \over \alpha}
+ B_0 \left({\alpha \over \pi}\right)^2
- {3\alpha^3 \over 2\pi}ln^2 {1\over \alpha} + {\cal O}(\alpha^3 ln\alpha)
\right]
\nonumber \\
&\simeq& (7.0382 + 3.9\times 10^{-5}B_0)\mu s^{-1}.
\label{xxx}
\end{eqnarray}
The $B_0$ term
parametrizes the non-logarithmic two-loop effects which have yet
to be calculated.

On the experimental side,
there have been a number of measurements
of the lifetime of orthopositronium. The most accurate
measurements are given in the table below:
\vskip 1.0cm

{\begin{center}
\begin{tabular}{|l|l|l|l|}
\hline
Reference$\;\;\;\;\;\;\;$
&$\Gamma_{oPs} (\mu s^{-1})$ $\;\;\;$ 
&Method$\;\;\;\;\;\;\;\;\;\;\;\;$
&$\Gamma_{coll}$$\;\;\;\;\;\;\;\;\;\;\;\;$\\
\hline
Ann Arbor\cite{aa1}&$7.0482\pm 0.0016$&Vacuum Cavity&$\sim (3-10)\Gamma_{oPs}$\\
Ann Arbor\cite{aa2}&$7.0514\pm 0.0014$&Gas&$\sim 10^3 \Gamma_{oPs}$\\
Tokyo\cite{tok}&$7.0398\pm 0.0029$&Powder&$\sim 10^4 \Gamma_{oPs}$\\
\hline
\end{tabular}\end{center}}

Table Caption:
Some measurements of the orthopositronium lifetime.
The last column is an estimate of the mean scattering
length of the orthopositronium in the experiment.

\vskip 1cm

Comparison of the theoretical prediction with the experimental
results suggests a statistically significant discrepancy
between the Theory/Tokyo results and the Ann Arbour results.\
(Note that the $B_0$ term would have to be anomalously large,
$\sim 300$, 
in order to make the theory and
the Ann Arbour result agree).
Originally it was proposed that an exotic o-Ps decay mode would solve 
the Theory/Ann Arbor discrepancy.\ The puzzle seemed so 
intriguing that even the strongly forbidden 
decay o-Ps $\rightarrow 2 \gamma$ was searched for as a candidate for 
solution.\ Now, it is believed that practically all  
possible contributions to the o-Ps decay rate from  non-standard  
annihilation modes are  excluded experimentally, 
for a review see \cite{review} ( and also \cite{tok}) and references 
therein.\
 The purpose of this letter is to see if the oscillations
of orthopositronium with its mirror analogue can resolve
this discrepancy.

Photon - mirror photon kinetic mixing
is described by the interaction Lagrangian density
\begin{equation}
{\cal L} = \epsilon F^{\mu \nu} F'_{\mu \nu},
\label{ek}
\end{equation}
where $F^{\mu \nu}$ ($F'_{\mu \nu}$) is the field strength 
tensor for electromagnetism (mirror electromagnetism).
This type of Lagrangian term is gauge invariant 
and renormalizable and can exist at tree level\cite{fh,flv}
or maybe induced radiatively in models without $U(1)$ 
gauge symmetries (such as grand unified theories)\cite{bob,gl,cf}.
The effect of ordinary photon - mirror photon kinetic mixing
is to give the mirror charged particles a small electric
charge\cite{bob,gl,flv}. That is, they couple to ordinary photons with
charge $2\epsilon e$\footnote{Note that the direct experimental
bound on $\epsilon$ from searches for `milli-charged' particles
is $\epsilon \stackrel{<}{\sim} 10^{-5}$\cite{slac}.}.

Orthopositronium is connected via a one-photon
annihilation diagram to its mirror version (o-Ps')\cite{gl}.
This breaks the degeneracy between o-Ps and o-Ps' so that
the vacuum energy eigenstates are (o-Ps + o-Ps')$/\sqrt{2}$ and 
(o-Ps - o-Ps')$/\sqrt{2}$,
which are split in energy by $\Delta E = 2h\epsilon f$, 
where $f = 8.7\times 10^4$ MHz is the contribution to the
ortho-para splitting from the one-photon annihilation diagram
involving o-Ps\cite{gl}.
Thus the interaction eigenstates are maximal combinations
of mass eigenstates which implies that o-Ps oscillates
into o-Ps' with probability: 
\begin{equation}
P({\rm o-Ps} \to {\rm o-Ps'}) = \sin^2 \omega t, 
\end{equation}
where $\omega = 2\pi\epsilon f$.
Note that the probability $P({\rm o-Ps} \to {\rm o-Ps'})$  
can, in principle, also be affected by an 
additional splitting of o-Ps and o-Ps' states 
by an external electric or magnetic  field\cite{gnin}.\

Let us first review the simplest case of o-Ps $\to$ o-Ps' 
oscillations in vacuum\cite{gl}. In this case,
because the mirror decays are not detected,
this leads to an {\it apparent} increase in the decay
rate, since the number of o-Ps, $N$ satisfies
\begin{equation}
N = \cos^2 wt e^{-\Gamma^{sm} t}
\simeq exp [-t(\Gamma^{sm} + w^2t)],
\end{equation}
where
$\Gamma^{sm}$ is the standard model decay rate of o-Ps (i.e. when
the oscillation length goes to infinity).
Thus $\Gamma^{eff} \approx \Gamma^{sm} + w^2/\Gamma^{sm}$.

Actually the above computation is not applicable to
any of the experiments listed in the table because in each case
the collision rate is larger than the decay rate and
loss of coherence due to the collisions must be included\cite{gnin}.\

Observe that there is the interesting possibility that
two different experiments could
get different values for the lifetime because of
the different collision rates of the orthopositronium.
It is well known that collisions damp the oscillations
and in the limit where the collision rate is much larger
than the decay rate (or oscillation frequency; whichever is smaller)
the effect of the oscillations becomes negligible (Quantum Zeno effect).
Let us assume for now that the theory computation is accurate
(i.e. the unknown quantity $B_0$ is not anomolously large). The agreement
with the Tokyo experiment can be explained because of the
very large collision rate of the orthopositronium in the powder.
However because of the two different collision rates of the 
two Ann Arbour experiments, they cannot both be explained.
Interestingly, some doubts about the Ann Arbour gas experiment
have emerged due to the fact that the thermalization time of o-Ps
(time of slowing of the orthopositronium atoms by collisions in matter)
needs to be taken into account  for a
precision measurement of the o-Ps decay rate, see e.g. \cite{tok},
\cite{chang}, \cite{asai}.\  If the
o-Ps atoms are not completely thermalized, they  
 are moving faster and their decay rate is spuriously higher
because of the higher pickoff annihilation rate.\
 This could
result in a systematic error in the determination of $\Gamma_{oPs}$.\
A similar conclusion has been drawn
  in a recent paper of Skalsey et al.
\cite{skalsey}, where it is claimed that at the lowest pressures
used in the gas measurements \cite{aa2} o-Ps were indeed not completely 
thermalized.\ This is crutial, since the Ann Arbor team used 
the asymptotic o-Ps decay rate extrapolated from low-pressure measurements.\  
The size of the correction on the o-Ps decay rate due to 
incomplete thermalization still has to be determined.\ 

If we ignore this gas result then
the discrepancy between the theory/Tokyo results and the
Ann Arbour vacuum cavity experiment can be explained by
the orthopositronium-mirror orthopositronium oscillation
mechanism. To see how this works in detail we need
to consider the case where
$\Gamma_{coll} \gg \Gamma^{sm}$, then the evolution
of the number of orthopositronium states, $N$, satisfies:
\begin{equation}
{dN \over dt} \simeq
- \Gamma^{sm} N - \Gamma_{coll}N\rho,
\label{1}
\end{equation}
where the second term is the rate at which o-Ps oscillates into o-Ps'
(whose subsequent decays are not detected). In this term,
$\rho$ denotes the average of oscillation probability
over the collision
time. That is,
\begin{equation}
\rho
\equiv \Gamma_{coll}\int^t_{0} e^{-\Gamma_{coll}t'} \sin^2 wt' dt' \simeq
\Gamma_{coll}\int^t_0 e^{-\Gamma_{coll}t'} (wt')^2 dt',
\end{equation}
where we have used the constraint
that the oscillation probability is small, i.e. $wt \ll 1$
(which must be the case given that the discrepancy between
say the vacuum cavity experiment and Tokyo experiment is less than 
1 percent).
So long as $t \gg 1/\Gamma_{coll}$ (which is an excellent
approximation for the gas and powder experiments and a reasonable
one for the vacuum cavity experiment) then
\begin{equation}
\rho \simeq {2w^2\over \Gamma_{coll}^2 }.
\end{equation} 
Thus substituting the above equation into Eq.(\ref{1})
we have
\begin{equation}
\Gamma^{eff} \simeq \Gamma^{sm} + {2w^2 \over \Gamma_{coll}}
= \Gamma^{sm}\left(1 + {2w^2 \over \Gamma_{coll}\Gamma^{sm}}\right). 
\end{equation}
Thus the high decay rate measured in the vacuum cavity experiment
relative to the theory result 
can be explained provied that
\begin{equation}
{2w^2 \over \Gamma_{coll}\Gamma^{sm}} \simeq 0.0014\pm 0.0002.
\end{equation} 
The experiment involves a range of cavity sizes
where the decay rate is obtained by extrapolation to
an infinite volume. From Ref.\cite{aa1} we estimate
that the largest cavity corresponds to $\Gamma_{coll} \sim 3\Gamma_{oPs}$,
which, neglecting the contribution from external fields 
(which are in fact negligible in this case\cite{gnin}),
implies that
\begin{equation}
w^2 \sim 2\times 10^{-3}\Gamma^2_{oPs} \Rightarrow
\epsilon \simeq (5\pm 1)\times 10^{-7}. 
\label{ep}
\end{equation}
The much larger collision rates
of the Tokyo (and gas) experiments means that the oscillations can have no 
effect on these experiments. Thus the high value of the vacuum
experiment relative to the 
theory/Tokyo results can be explained. However the high value of 
the Michigan gas experiment cannot be explained by oscillations but is 
presumably due to larger than expected systematic errors (as
discussed earlier). 

Interestingly the value of $\epsilon$ identified in Eq.(\ref{ep})
is consistent with all
known experimental and cosmological bounds (including SN1987)
with the exception of the big bang nucleosynthesis bound (see 
figure 1 of Ref.\cite{raf} for a review and references).
Big bang nucleosynthesis suggests the bound\cite{cg} 
$\epsilon \stackrel{<}{\sim} 3\times 10^{-8}$.
Thus, if it were experimentally proven that $\epsilon$ is
as large as Eq.(\ref{ep}) then it would presumably mean that
some of the assumptions of big bang nucleosynthesis would
have to be modified.

The experimental signature of  the o-Ps $\rightarrow$ o-Ps'  
oscillations 
is the 'disappearance' of an energy deposition of $\sim$ 1 MeV,
which is expected from the ordinary o-Ps annihilation
in a 4$\pi$ calorimeter
surrounding the o-Ps formation region, i.e. an invisible decay of o-Ps.\
The first experiment on such decays was performed  
a long time ago \cite{atojan}, and then repeated 
with higher sensitivity  \cite{mitsui}.\ The results exclude
 contributions to the o-Ps decay rate from 
invisible decay modes (such as o-Ps$\rightarrow \nu \nu$, millicharged 
particles, etc..) at the level of BR(o-Ps$\rightarrow 
invisible) <3\cdot 10^{-6}$, 
but are not very sensitive to the o-Ps $\to$ o-Ps' oscillation
mechanism because of the high collision rate in these experiments. 
Indeed the limit on $\epsilon$  extracted from the results of 
ref. \cite{mitsui}, taking into account the 
suppression collision factor, is $\epsilon < 10^{-6}$ \cite{gnin} and is not 
strong enough to exclude a possible mirror contribution given by Eqs.(8,10) to
the o-Ps decay rate.\ Thus,
a vacuum experiment with significantly higher sensitivity will be necessary 
to confirm or rule out the mirror world effect.\ The best approach
would be to combine the
lifetime and invisible decay rate measurements in a single cavity experiment.\
In this case one would have a good cross-check: 
the higher o-Ps decay rate the larger peak at zero energy.\

If the photon - mirror photon mixing is as
large as suggested 
here then there will be a number of interesting implications.
For example, any mirror matter in the center of the sun can become quite
hot due to the absorption of ordinary photons.
The subsequent radiation of mirror photons by the mirror matter
would be absorbed by the surrounding ordinary matter providing
a new source of energy transport in the solar interior.
Another implication of the large photon - mirror photon kinetic
mixing is that it may be large enough (depending on
the chemical composition of the mirror planet) to make mirror planets
opaque to ordinary photons. Thus the recent observations\cite{ob} of
a transit of the extrasolar planet HD 209458
does not exclude the hypothesis\cite{hyp} that the close-in extrasolar
planets are mirror planets. The remaining prediction
of the mirror extrasolar planet hypothesis is that they cannot
reflect the light from the star (i.e. their albedo is 
essentially zero). This prediction is currently being tested\cite{test}.
The photon - mirror photon kinetic mixing may 
provide a useful way of observing a mirror supernova explosion.
This is because the rapid conversion of mirror matter into
ordinary matter (via $e'^+ e'^- \to e^+ e^- \to \gamma + \gamma$)
may provide a burst of ordinary light. Indeed it may be possible 
that this is the mechanism which can realize 
Blinnikov's proposal\cite{bl} that
gamma ray bursts are infact collapsing (or merging) mirror stars.
Finally, it is interesting to note that
composites of mirror
matter with no net mirror charge will exert small Van der Waals forces 
upon ordinary matter, possibly making terrestrial encounters with cosmic 
chunks of mirror matter also observable \cite{gl}, \cite{der}.\
Indeed the collision of a mirror object (e.g. mirror
asteroid) with the earth may well
be a health hazard as the mirror object will not burn
up in the atmosphere, but may nevertheless deposit a 
large amount of energy when it hits the surface as the
object losses energy due to electromagnetic interactions
induced by the kinetic mixing. Such an impact would not be
expected to leave any significant crater. \footnote{
It is of course amuzing to speculate that the 1908 Tunguska event 
may have been one such event since no impact crater
was ever found.}. 
 
While it is fun to speculate about the effects of large
photon - mirror photon kinetic mixing, real progress
requires experiments.
Given the already strong evidence for the mirror world
coming from dark matter\cite{dark} and the neutrino physics 
anomalies\cite{neu} (as well as the intuitive expectation
that nature should be left-right symmetric)
it is obviously important to experimentally determine
whether the orthopositronium lifetime puzzle is one
more window on the mirror world. 
We hope that future experiments will shed light on this issue \cite{mesh}.

\vskip 0.8cm
\noindent
{\bf Note Added}
\vskip 0.5cm
After completion of this paper the preprint\cite{new} appeared
which calculates the unknown $B_0$
term in Eq.(\ref{xxx}). They obtain the value
$B_0 \simeq 44$ which gives a theoretical orthopositronium decay rate of
$\Gamma = 7.0399\ \mu s^{-1}$\cite{new}.
Thus, the theoretical calculation agrees with the Tokyo experiment
which we in fact anticipated in this paper.
The theoretical calculation is
about 5 standard deviations below the vacuum cavity experiment,
which in our paper is explained by the influence of the 
mirror universe.

\newpage

\vspace{0.5cm}
{\bf Asknowledgments}

\vskip 0.5cm

S.G. would like to thank S. Glashow for stimulating interest and 
valuable discussions and S. Asai for communications on 
o-Ps thermalization.
R.F. thanks S. Blinnikov, Z. Silagadze and R. Volkas 
for discussions/correspondence. R.F. is an Australian Research Fellow.

\end{document}